\documentclass[conference,a4paper]{IEEEtran}
\IEEEoverridecommandlockouts
\usepackage[acronyms,nonumberlist,nopostdot,nomain,nogroupskip,acronymlists={hidden}]{glossaries}
\newglossary[algh]{hidden}{acrh}{acnh}{Hidden Acronyms}
\usepackage{silence}
\usepackage{amsmath,amssymb,amsfonts}
\usepackage{algorithmic}
\usepackage{textcomp}
\usepackage{courier} 
\usepackage{xcolor}
\usepackage{booktabs}
\usepackage{makecell}
\usepackage{balance}
\usepackage{url}

\usepackage{tikz}
\usepackage{placeins}  % Aggiungi nel preambolo
\usepackage{pgfplots}
\pgfplotsset{compat=newest}
\pgfplotsset{plot coordinates/math parser=false}
\newlength\fheight
\newlength\fwidth
\usetikzlibrary{plotmarks,patterns,decorations.pathreplacing,backgrounds,calc,arrows,arrows.meta,spy,matrix,scopes}
\usepgfplotslibrary{patchplots,groupplots}
\usepackage{tikzscale}
\pgfplotsset{compat=1.18}

\def\BibTeX{{\rm B\kern-.05em{\sc i\kern-.025em b}\kern-.08em
    T\kern-.1667em\lower.7ex\hbox{E}\kern-.125emX}}

\expandafter\def\csname ver@fixltx2e.sty\endcsname{}

\newif\ifexttikz
\exttikzfalse
\ifexttikz
	\usetikzlibrary{external}
	\usepackage{fontspec}
\fi

\newif\ifexttikz
\exttikzfalse
\ifexttikz
\else
\usepackage{tikzpagenodes,etoolbox}
\usetikzlibrary{calc}
\usepackage[contents={}]{background}
\AddEverypageHook{%
\ifnumequal{\thepage}{1}{%
    \tikz[remember picture,overlay]{%
        \node[draw,
        minimum width=1.03\textwidth,
        text width=1.02\textwidth,
        font=\footnotesize
        ]
        at ($(current page header area) - (0,5pt)$)
        {%
        F. Olimpieri, N. Giustini, A. Lacava, S. D'Oro, T. Melodia and F. Cuomo, "LibIQ: Toward Real-Time Spectrum Classification in O-RAN dApps," 2025 23rd Mediterranean Communication and Computer Networking Conference (MedComNet), Cagliari, Italy, 2025, pp. 1-6, doi: 10.1109/MedComNet65822.2025.11100289-. 
        };
        \node[draw,
        minimum width=1.03\textwidth,
        text width=1.02\textwidth,
        font=\footnotesize
        ]
        at (current page footer area)
        {%
        ©2025 IEEE. Personal use of this material is permitted. Permission from IEEE must be obtained for all other uses, in any current or future media, including reprinting/republishing this material for advertising or promotional purposes, creating new collective works, for resale or redistribution to servers or lists, or reuse of any copyrighted component of this work in other works.
        };
    }%
}{}%end ifnumequal
}
\fi
\usepackage[font=footnotesize]{caption}
\usepackage{subcaption}
\newacronym{3gpp}{3GPP}{3rd Generation Partnership Project}
\newacronym{4g}{4G}{4th generation}
\newacronym{5g}{5G}{5th generation}
\newacronym{6g}{6G}{6th generation}
\newacronym{5gc}{5GC}{5G Core}
\newacronym{adc}{ADC}{Analog to Digital Converter}
\newacronym{aerpaw}{AERPAW}{Aerial Experimentation and Research Platform for Advanced Wireless}
\newacronym{ai}{AI}{Artificial Intelligence}
\newacronym{aimd}{AIMD}{Additive Increase Multiplicative Decrease}
\newacronym{am}{AM}{Acknowledged Mode}
\newacronym{amc}{AMC}{Adaptive Modulation and Coding}
\newacronym{amf}{AMF}{Access and Mobility Management Function}
\newacronym{aops}{AOPS}{Adaptive Order Prediction Scheduling}
\newacronym{api}{API}{Application Programming Interface}
\newacronym{apn}{APN}{Access Point Name}
\newacronym{ap}{AP}{Application Protocol}
\newacronym{aqm}{AQM}{Active Queue Management}
\newacronym{ausf}{AUSF}{Authentication Server Function}
\newacronym{avc}{AVC}{Advanced Video Coding}
\newacronym{awgn}{AGWN}{Additive White Gaussian Noise}
\newacronym{balia}{BALIA}{Balanced Link Adaptation Algorithm}
\newacronym{bbu}{BBU}{Base Band Unit}
\newacronym{bdp}{BDP}{Bandwidth-Delay Product}
\newacronym{ber}{BER}{Bit Error Rate}
\newacronym{bf}{BF}{Beamforming}
\newacronym{bler}{BLER}{Block Error Rate}
\newacronym{brr}{BRR}{Bayesian Ridge Regressor}
\newacronym{bs}{BS}{Base Station}
\newacronym{bsr}{BSR}{Buffer Status Report}
\newacronym{bss}{BSS}{Business Support System}
\newacronym{ca}{CA}{Carrier Aggregation}
\newacronym{caas}{CaaS}{Connectivity-as-a-Service}
\newacronym{cb}{CB}{Code Block}
\newacronym{cc}{CC}{Congestion Control}
\newacronym{ccid}{CCID}{Congestion Control ID}
\newacronym{cco}{CC}{Carrier Component}
\newacronym{cdd}{CDD}{Cyclic Delay Diversity}
\newacronym{cdf}{CDF}{Cumulative Distribution Function}
\newacronym{cdn}{CDN}{Content Distribution Network}
\newacronym{cir}{CIR}{Channel Impulse Response}
\newacronym{cli}{CLI}{Command-line Interface}
\newacronym{cn}{CN}{Core Network}
\newacronym{cnn}{CNN}{Convolutional Neural Network}
\newacronym{codel}{CoDel}{Controlled Delay Management}
\newacronym{comac}{COMAC}{Converged Multi-Access and Core}
\newacronym{cord}{CORD}{Central Office Re-architected as a Datacenter}
\newacronym{cornet}{CORNET}{COgnitive Radio NETwork}
\newacronym{cosmos}{COSMOS}{Cloud Enhanced Open Software Defined Mobile Wireless Testbed for City-Scale Deployment}
\newacronym{cots}{COTS}{Commercial Off-the-Shelf}
\newacronym{cp}{CP}{Control Plane}
\newacronym{cyp}{CP}{Cyclic Prefix}
\newacronym{up}{UP}{User Plane}
\newacronym{cpu}{CPU}{Central Processing Unit}
\newacronym{cqi}{CQI}{Channel Quality Information}
\newacronym{cql}{CQL}{Conservative Q-Learning}
\newacronym{cr}{CR}{Cognitive Radio}
\newacronym{cran}{CRAN}{Cloud \gls{ran}}
\newacronym{crs}{CRS}{Cell Reference Signal}
\newacronym{csi}{CSI}{Channel State Information}
\newacronym{csirs}{CSI-RS}{Channel State Information - Reference Signal}
\newacronym{csv}{CSV}{Comma Separated Values}
\newacronym{cu}{CU}{Central Unit}
\newacronym{cucp}{CU-CP}{Central Unit Control Plane}
\newacronym{cuup}{CU-UP}{Central Unit User Plane}
\newacronym{d2tcp}{D$^2$TCP}{Deadline-aware Data center TCP}
\newacronym{d3}{D$^3$}{Deadline-Driven Delivery}
\newacronym{dac}{DAC}{Digital to Analog Converter}
\newacronym{dag}{DAG}{Directed Acyclic Graph}
\newacronym{dapp}{DApp}{Distributed Application}
\newacronym{dapps}{DApps}{Distributed Applications}
\newacronym{das}{DAS}{Distributed Antenna System}
\newacronym{dash}{DASH}{Dynamic Adaptive Streaming over HTTP}
\newacronym{dbscan}{DBSCAN}{Density-Based Spatial
Clustering of Applications with Noise}
\newacronym{dc}{DC}{Dual Connectivity}
\newacronym{dccp}{DCCP}{Datagram Congestion Control Protocol}
\newacronym{dce}{DCE}{Direct Code Execution}
\newacronym{dci}{DCI}{Downlink Control Information}
\newacronym{dctcp}{DCTCP}{Data Center TCP}
\newacronym{dl}{DL}{Downlink}
\newacronym{dmr}{DMR}{Deadline Miss Ratio}
\newacronym{dmrs}{DMRS}{DeModulation Reference Signal}
\newacronym{dqn}{DQN}{Deep Q-Network}
\newacronym{dtw}{DTW}{Dynamic Time Warping}
\newacronym{drlcc}{DRL-CC}{Deep Reinforcement Learning Congestion Control}
\newacronym{drs}{DRS}{Discovery Reference Signal}
\newacronym{du}{DU}{Distributed Unit}
\newacronym{ee}{EE}{Energy Efficiency}
\newacronym{e2e}{E2E}{end-to-end}
\newacronym{earfcn}{EARFCN}{E-UTRA Absolute Radio Frequency Channel Number}
\newacronym{ecaas}{ECaaS}{Edge-Cloud-as-a-Service}
\newacronym{ecn}{ECN}{Explicit Congestion Notification}
\newacronym{edf}{EDF}{Earliest Deadline First}
\newacronym{embb}{eMBB}{Enhanced Mobile Broadband}
\newacronym{empower}{EMPOWER}{EMpowering transatlantic PlatfOrms for advanced WirEless Research}
\newacronym{enb}{eNB}{evolved Node Base}
\newacronym{endc}{EN-DC}{E-UTRAN-\gls{nr} \gls{dc}}
\newacronym{epc}{EPC}{Evolved Packet Core}
\newacronym{eps}{EPS}{Evolved Packet System}
\newacronym{es}{ES}{Edge Server}
\newacronym{etsi}{ETSI}{European Telecommunications Standards Institute}
\newacronym[firstplural=Estimated Times of Arrival (ETAs)]{eta}{ETA}{Estimated Time of Arrival}
\newacronym{eutran}{E-UTRAN}{Evolved Universal Terrestrial Access Network}
\newacronym{faas}{FaaS}{Function-as-a-Service}
\newacronym{fapi}{FAPI}{Functional Application Platform Interface}
\newacronym{fdd}{FDD}{Frequency Division Duplexing}
\newacronym{fdm}{FDM}{Frequency Division Multiplexing}
\newacronym{fdma}{FDMA}{Frequency Division Multiple Access}
\newacronym{fed4fire}{FED4FIRE+}{Federation 4 Future Internet Research and Experimentation Plus}
\newacronym{fir}{FIR}{Finite Impulse Response}
\newacronym{fit}{FIT}{Future \acrlong{iot}}
\newacronym{fpga}{FPGA}{Field Programmable Gate Array}
\newacronym{fr1}{FR1}{Frequency Range 1}
\newacronym{fr2}{FR2}{Frequency Range 2}
\newacronym{fs}{FS}{Fast Switching}
\newacronym{fscc}{FSCC}{Flow Sharing Congestion Control}
\newacronym{ftp}{FTP}{File Transfer Protocol}
\newacronym{fw}{FW}{Flow Window}
\newacronym{ge}{GE}{Gaussian Elimination}
\newacronym{gnb}{gNB}{Next Generation Node Base}
\newacronym{gop}{GOP}{Group of Pictures}
\newacronym{gpr}{GPR}{Gaussian Process Regressor}
\newacronym{gpu}{GPU}{Graphics Processing Unit}
\newacronym{gtp}{GTP}{GPRS Tunneling Protocol}
\newacronym{gan}{GAN}{Generative Adversarial Network}
\newacronym{gui}{GUI}{Graphical User Interface}
\newacronym{gtpc}{GTP-C}{GPRS Tunnelling Protocol Control Plane}
\newacronym{gtpu}{GTP-U}{GPRS Tunnelling Protocol User Plane}
\newacronym{gtpv2c}{GTPv2-C}{\gls{gtp} v2 - Control}
\newacronym{gw}{GW}{Gateway}
\newacronym{harq}{HARQ}{Hybrid Automatic Repeat reQuest}
\newacronym{hetnet}{HetNet}{Heterogeneous Network}
\newacronym{hh}{HH}{Hard Handover}
\newacronym{hol}{HOL}{Head-of-Line}
\newacronym{hqf}{HQF}{Highest-quality-first}
\newacronym{hss}{HSS}{Home Subscription Server}
\newacronym{http}{HTTP}{HyperText Transfer Protocol}
\newacronym{ia}{IA}{Initial Access}
\newacronym{iab}{IAB}{Integrated Access and Backhaul}
\newacronym{ic}{IC}{Incident Command}
\newacronym{ietf}{IETF}{Internet Engineering Task Force}
\newacronym{imsi}{IMSI}{International Mobile Subscriber Identity}
\newacronym{imt}{IMT}{International Mobile Telecommunication}
\newacronym{iot}{IoT}{Internet of Things}
\newacronym{ip}{IP}{Internet Protocol}
\newacronym{ipc}{IPC}{inter-process communication}
\newacronym{itu}{ITU}{International Telecommunication Union}
\newacronym{knn}{KNN}{k-nearest neighbors}
\newacronym{kpi}{KPI}{Key Performance Indicator}
\newacronym{kpm}{KPM}{Key Performance Measurement}
\newacronym{kvm}{KVM}{Kernel-based Virtual Machine}
\newacronym{los}{LOS}{Line-of-Sight}
\newacronym{lsm}{LSM}{Link-to-System Mapping}
\newacronym{lstm}{LSTM}{Long Short Term Memory}
\newacronym{lte}{LTE}{Long Term Evolution}
\newacronym{lxc}{LXC}{Linux Container}
\newacronym{m2m}{M2M}{Machine to Machine}
\newacronym{mac}{MAC}{Medium Access Control}
\newacronym{manet}{MANET}{Mobile Ad Hoc Network}
\newacronym{mano}{MANO}{Management and Orchestration}
\newacronym{mc}{MC}{Multi-Connectivity}
\newacronym{mcc}{MCC}{Mobile Cloud Computing}
\newacronym{mchem}{MCHEM}{Massive Channel Emulator}
\newacronym{mcs}{MCS}{Modulation and Coding Scheme}
\newacronym{mec}{MEC}{Multi-access Edge Computing}
\newacronym{mec2}{MEC}{Mobile Edge Cloud}
\newacronym{mfc}{MFC}{Mobile Fog Computing}
\newacronym{mgen}{MGEN}{Multi-Generator}
\newacronym{mi}{MI}{Mutual Information}
\newacronym{mib}{MIB}{Master Information Block}
\newacronym{miesm}{MIESM}{Mutual Information Based Effective SINR}
\newacronym{mimo}{MIMO}{Multiple Input, Multiple Output}
\newacronym{ml}{ML}{Machine Learning}
\newacronym{mlr}{MLR}{Maximum-local-rate}
\newacronym[plural=\gls{mme}s,firstplural=Mobility Management Entities (MMEs)]{mme}{MME}{Mobility Management Entity}
\newacronym{mmtc}{mMTC}{Massive Machine-Type Communications}
\newacronym{mmwave}{mmWave}{millimeter wave}
\newacronym{mpdccp}{MP-DCCP}{Multipath Datagram Congestion Control Protocol}
\newacronym{mptcp}{MPTCP}{Multipath TCP}
\newacronym{mr}{MR}{Maximum Rate}
\newacronym{mrdc}{MR-DC}{Multi \gls{rat} \gls{dc}}
\newacronym{mse}{MSE}{Mean Square Error}
\newacronym{mss}{MSS}{Maximum Segment Size}
\newacronym{mt}{MT}{Mobile Termination}
\newacronym{mtd}{MTD}{Machine-Type Device}
\newacronym{mtu}{MTU}{Maximum Transmission Unit}
\newacronym{mumimo}{MU-MIMO}{Multi-user \gls{mimo}}
\newacronym{mvno}{MVNO}{Mobile Virtual Network Operator}
\newacronym{nalu}{NALU}{Network Abstraction Layer Unit}
\newacronym{nas}{NAS}{Network Attached Storage}
\newacronym{nat}{NAT}{Network Address Translation}
\newacronym{nbiot}{NB-IoT}{Narrow Band IoT}
\newacronym{nfv}{NFV}{Network Function Virtualization}
\newacronym{nfvi}{NFVI}{Network Function Virtualization Infrastructure}
\newacronym{ni}{NI}{Network Interfaces}
\newacronym{nic}{NIC}{Network Interface Card}
\newacronym{nlos}{NLOS}{Non-Line-of-Sight}
\newacronym{now}{NOW}{Non Overlapping Window}
\newacronym{nsm}{NSM}{Network Service Mesh}
\newacronym[type=hidden]{nr}{NR}{New Radio}
\newacronym{nextg}{NextG}{Next Generation}
\newacronym{nrf}{NRF}{Network Repository Function}
\newacronym{nsa}{NSA}{Non Stand Alone}
\newacronym{nse}{NSE}{Network Slicing Engine}
\newacronym{nssf}{NSSF}{Network Slice Selection Function}
\newacronym{ngrg}{nGRG}{next Generation Research Group}
\newacronym{o2i}{O2I}{Outdoor to Indoor}
\newacronym{oai}{OAI}{OpenAirInterface}
\newacronym{oaicn}{OAI-CN}{\gls{oai} \acrlong{cn}}
\newacronym{oairan}{OAI-RAN}{\acrlong{oai} \acrlong{ran}}
\newacronym{oam}{OAM}{Operations, Administration and Maintenance}
\newacronym{ofdm}{OFDM}{Orthogonal Frequency Division Multiplexing}
\newacronym{olia}{OLIA}{Opportunistic Linked Increase Algorithm}
\newacronym{omec}{OMEC}{Open Mobile Evolved Core}
\newacronym{onap}{ONAP}{Open Network Automation Platform}
\newacronym{onf}{ONF}{Open Networking Foundation}
\newacronym{onos}{ONOS}{Open Networking Operating System}
\newacronym{oom}{OOM}{\gls{onap} Operations Manager}
\newacronym{opnfv}{OPNFV}{Open Platform for \gls{nfv}}
\newacronym{orbit}{ORBIT}{Open-Access Research Testbed for Next-Generation Wireless Networks}
\newacronym{os}{OS}{Operating System}
\newacronym{osm2}{OSM}{Open Street Map}
\newacronym{oss}{OSS}{Operations Support System}
\newacronym{ota}{OTA}{Over-The-Air}
\newacronym{pa}{PA}{Position-aware}
\newacronym{pase}{PASE}{Prioritization, Arbitration, and Self-adjusting Endpoints}
\newacronym{pawr}{PAWR}{Platforms for Advanced Wireless Research}
\newacronym{pbch}{PBCH}{Physical Broadcast Channel}
\newacronym{pca}{PCA}{Principal Component Analysis}
\newacronym{psd}{PSD}{Power Spectral Density}
\newacronym{pcef}{PCEF}{Policy and Charging Enforcement Function}
\newacronym{pcfich}{PCFICH}{Physical Control Format Indicator Channel}
\newacronym{pcrf}{PCRF}{Policy and Charging Rules Function}
\newacronym{pdcch}{PDCCH}{Physical Downlink Control Channel}
\newacronym{pdcp}{PDCP}{Packet Data Convergence Protocol}
\newacronym{pdsch}{PDSCH}{Physical Downlink Shared Channel}
\newacronym{pdu}{PDU}{Packet Data Unit}
\newacronym{pf}{PF}{Proportional Fair}
\newacronym{pgw}{PGW}{Packet Gateway}
\newacronym{phich}{PHICH}{Physical Hybrid ARQ Indicator Channel}
\newacronym{phy}{PHY}{Physical}
\newacronym{pl}{PL}{Path Loss}
\newacronym{pmch}{PMCH}{Physical Multicast Channel}
\newacronym{pmi}{PMI}{Precoding Matrix Indicators}
\newacronym{powder}{POWDER}{Platform for Open Wireless Data-driven Experimental Research}
\newacronym{ppo}{PPO}{Proximal Policy Optimization}
\newacronym{ppp}{PPP}{Poisson Point Process}
\newacronym{prach}{PRACH}{Physical Random Access Channel}
\newacronym{prb}{PRB}{Physical Resource Block}
\newacronym{psnr}{PSNR}{Peak Signal to Noise Ratio}
\newacronym{pss}{PSS}{Primary Synchronization Signal}
\newacronym{pucch}{PUCCH}{Physical Uplink Control Channel}
\newacronym{pusch}{PUSCH}{Physical Uplink Shared Channel}
\newacronym{qam}{QAM}{Quadrature Amplitude Modulation}
\newacronym{qci}{QCI}{\gls{qos} Class Identifier}
\newacronym{qoe}{QoE}{Quality of Experience}
\newacronym{qos}{QoS}{Quality of Service}
\newacronym{quic}{QUIC}{Quick UDP Internet Connections}
\newacronym{rach}{RACH}{Random Access Channel}
\newacronym{ran}{RAN}{Radio Access Network}
\newacronym[firstplural=Radio Access Technologies (RATs)]{rat}{RAT}{Radio Access Technology}
\newacronym{rbg}{RBG}{Resource Block Group}
\newacronym{rcn}{RCN}{Research Coordination Network}
\newacronym{rc}{RC}{RAN Control}
\newacronym{rec}{REC}{Radio Edge Cloud}
\newacronym{red}{RED}{Random Early Detection}
\newacronym{renew}{RENEW}{Reconfigurable Eco-system for Next-generation End-to-end Wireless}
\newacronym{rf}{RF}{Radio Frequency}
\newacronym{rfi}{RFI}{Radio Frequency Interference}
\newacronym{rfc}{RFC}{Request for Comments}
\newacronym{rfr}{RFR}{Random Forest Regressor}
\newacronym{ric}{RIC}{RAN Intelligent Controller}
\newacronym{rlc}{RLC}{Radio Link Control}
\newacronym{rlf}{RLF}{Radio Link Failure}
\newacronym{rlnc}{RLNC}{Random Linear Network Coding}
\newacronym{rmr}{RMR}{RIC Message Router}
\newacronym{rmse}{RMSE}{Root Mean Squared Error}
\newacronym{rnis}{RNIS}{Radio Network Information Service}
\newacronym{rr}{RR}{Round Robin}
\newacronym{rrc}{RRC}{Radio Resource Control}
\newacronym{rrm}{RRM}{Radio Resource Management}
\newacronym{rru}{RRU}{Remote Radio Unit}
\newacronym{rs}{RS}{Remote Server}
\newacronym{rsrp}{RSRP}{Reference Signal Received Power}
\newacronym{rsrq}{RSRQ}{Reference Signal Received Quality}
\newacronym{rss}{RSS}{Received Signal Strength}
\newacronym{rssi}{RSSI}{Received Signal Strength Indicator}
\newacronym{rt}{RT}{Real-time}
\newacronym{rtt}{RTT}{Round Trip Time}
\newacronym{ru}{RU}{Radio Unit}
\newacronym{rw}{RW}{Receive Window}
\newacronym{rx}{RX}{Receiver}
\newacronym{s1ap}{S1AP}{S1 Application Protocol}
\newacronym{sa}{SA}{standalone}
\newacronym{sack}{SACK}{Selective Acknowledgment}
\newacronym{sap}{SAP}{Service Access Point}
\newacronym{sc2}{SC2}{Spectrum Collaboration Challenge}
\newacronym{scef}{SCEF}{Service Capability Exposure Function}
\newacronym{sch}{SCH}{Secondary Cell Handover}
\newacronym{scoot}{SCOOT}{Split Cycle Offset Optimization Technique}
\newacronym{sctp}{SCTP}{Stream Control Transmission Protocol}
\newacronym{sdap}{SDAP}{Service Data Adaptation Protocol}
\newacronym{sdk}{SDK}{Software Development Kit}
\newacronym{sdm}{SDM}{Space Division Multiplexing}
\newacronym{sdma}{SDMA}{Spatial Division Multiple Access}
\newacronym{sdn}{SDN}{Software-defined Networking}
\newacronym{sdr}{SDR}{Software-defined Radio}
\newacronym{seba}{SEBA}{SDN-Enabled Broadband Access}
\newacronym{sgsn}{SGSN}{Serving GPRS Support Node}
\newacronym{sgw}{SGW}{Service Gateway}
\newacronym{si}{SI}{Study Item}
\newacronym{sib}{SIB}{Secondary Information Block}
\newacronym{sinr}{SINR}{Signal to Interference plus Noise Ratio}
\newacronym{sip}{SIP}{Session Initiation Protocol}
\newacronym{siso}{SISO}{Single Input, Single Output}
\newacronym{sla}{SLA}{Service Level Agreement}
\newacronym{sm}{SM}{Service Model}
\newacronym{smf}{SMF}{Session Management Function}
\newacronym{smo}{SMO}{Service Management and Orchestration}
\newacronym{sms}{SMS}{Short Message Service}
\newacronym{smsgmsc}{SMS-GMSC}{\gls{sms}-Gateway}
\newacronym{snr}{SNR}{Signal-to-Noise-Ratio}
\newacronym{son}{SON}{Self-Organizing Network}
\newacronym{sptcp}{SPTCP}{Single Path TCP}
\newacronym{srb}{SRB}{Service Radio Bearer}
\newacronym{swig}{SWIG}{Simplified Wrapper and Interface Generator}
\newacronym{srn}{SRN}{Standard Radio Node}
\newacronym{srs}{SRS}{Sounding Reference Signal}
\newacronym{ss}{SS}{Synchronization Signal}
\newacronym{sss}{SSS}{Secondary Synchronization Signal}
\newacronym{st}{ST}{Spanning Tree}
\newacronym{svc}{SVC}{Scalable Video Coding}
\newacronym{tb}{TB}{Transport Block}
\newacronym{tcp}{TCP}{Transmission Control Protocol}
\newacronym{tdd}{TDD}{Time Division Duplexing}
\newacronym{tdl}{TDL}{Tapped Delay Line}
\newacronym{tdm}{TDM}{Time Division Multiplexing}
\newacronym{tdma}{TDMA}{Time Division Multiple Access}
\newacronym{tfl}{TfL}{Transport for London}
\newacronym{tfrc}{TFRC}{TCP-Friendly Rate Control}
\newacronym{tft}{TFT}{Traffic Flow Template}
\newacronym{tgen}{TGEN}{Traffic Generator}
\newacronym{tip}{TIP}{Telecom Infra Project}
\newacronym{tm}{TM}{Transparent Mode}
\newacronym{to}{TO}{Telco Operator}
\newacronym{tr}{TR}{Technical Report}
\newacronym{trp}{TRP}{Transmitter Receiver Pair}
\newacronym{ts}{TS}{Technical Specification}
\newacronym{tti}{TTI}{Transmission Time Interval}
\newacronym{ttt}{TTT}{Time-to-Trigger}
\newacronym{tx}{TX}{Transmitter}
\newacronym{uas}{UAS}{Unmanned Aerial System}
\newacronym{uav}{UAV}{Unmanned Aerial Vehicle}
\newacronym{udm}{UDM}{Unified Data Management}
\newacronym{udp}{UDP}{User Datagram Protocol}
\newacronym{udr}{UDR}{Unified Data Repository}
\newacronym{ue}{UE}{User Equipment}
\newacronym{uhd}{UHD}{\gls{usrp} Hardware Driver}
\newacronym{ul}{UL}{Uplink}
\newacronym{um}{UM}{Unacknowledged Mode}
\newacronym{umi}{UMi}{Urban Micro}
\newacronym{uml}{UML}{Unified Modeling Language}
\newacronym{upa}{UPA}{Uniform Planar Array}
\newacronym{upf}{UPF}{User Plane Function}
\newacronym{urllc}{URLLC}{Ultra Reliable and Low Latency Communications}
\newacronym{usa}{U.S.}{United States}
\newacronym{usim}{USIM}{Universal Subscriber Identity Module}
\newacronym{usrp}{USRP}{Universal Software Radio Peripheral}
\newacronym{utc}{UTC}{Urban Traffic Control}
\newacronym{vim}{VIM}{Virtualization Infrastructure Manager}
\newacronym{vm}{VM}{Virtual Machine}
\newacronym{vnf}{VNF}{Virtual Network Function}
\newacronym{volte}{VoLTE}{Voice over \gls{lte}}
\newacronym{voltha}{VOLTHA}{Virtual OLT HArdware Abstraction}
\newacronym{vr}{VR}{Virtual Reality}
\newacronym{vran}{vRAN}{Virtualized \gls{ran}}
\newacronym{vss}{VSS}{Video Streaming Server}
\newacronym{wbf}{WBF}{Wired Bias Function}
\newacronym{wf}{WF}{Waterfilling}
\newacronym{wg}{WG}{Working Group}
\newacronym{wi}{WI}{Wireless InSite}
\newacronym{wlan}{WLAN}{Wireless Local Area Network}
\newacronym{osm}{OSM}{Open Source \gls{nfv} Management and Orchestration}
\newacronym{pnf}{PNF}{Physical Network Function}
\newacronym{mtc}{MTC}{Machine-type Communications}
\newacronym{mns}{MnS}{Management Services}
\newacronym{ves}{VES}{\gls{vnf} Event Stream}
\newacronym{ei}{EI}{Enrichment Information}
\newacronym{fh}{FH}{Fronthaul}
\newacronym{fft}{FFT}{Fast Fourier Transform}
\newacronym{laa}{LAA}{Licensed-Assisted Access}
\newacronym{plfs}{PLFS}{Physical Layer Frequency Signals}
\newacronym{ptp}{PTP}{Precision Time Protocol}
\newacronym{cbrs}{CBRS}{Citizen Broadband Radio Service}
\newacronym{otic}{OTIC}{Open Testing and Integration Center}
\newacronym{sba}{SBA}{Service-Based Architecture}
\newacronym{cif}{CI}{cyberinfrastructure}
\newacronym{sonic}{SONiC}{Software for Open Networking in the Cloud}
\newacronym{ocp}{OCP}{Open Compute Project}
\newacronym{snmp}{SNMP}{Simple Network Management Protocol}
\newacronym{raid}{RAID}{redundant array of independent disks}
\newacronym{nfs}{NFS}{Network File Storage}
\newacronym{ci}{CI}{Continuous Integration}
\newacronym{cd}{CD}{Continuous Deployment}
\newacronym{dtn}{DTN}{Data Transfer Node}

\newacronym{dns}{DNS}{Domain Name Service}
\newacronym{nrpe}{NRPE}{Nagios Remote Plugin Executor}
\newacronym{ldap}{LDAP}{Lightweight Directory Access Protocol}
\newacronym{lan}{LAN}{Local Area Network}
\newacronym{vlan}{VLAN}{Virtual LAN}

\newacronym{ipmi}{IPMI}{Intelligent Platform Management Interface}
\newacronym{tor}{ToR}{Top-of-the-Rack}
\newacronym{lmn}{LMN}{Large Memory Node}
\newacronym{bgp}{BGP}{Border Gateway Protocol}
\newacronym{dhcp}{DHCP}{Dynamic Host Configuration Protocol}
\newacronym{vrf}{VRF}{Virtual Routing and Forwarding}
\newacronym{vpn}{VPN}{Virtual Private Network}
\newacronym{rma}{RMA}{Return Merchandise Authorization}
\newacronym{hpc}{HPC}{High Performance Compute}

\newacronym{nu}{NU}{Northeastern University}
\newacronym{asic}{ASIC}{Application-specific Integrated Circuit}
\newacronym{rdma}{RDMA}{Remote Direct Memory Access}
\newacronym{roce}{RoCE}{RDMA over Converged Ethernet}
\newacronym{ovs}{OVS}{Open vSwitch}
\newacronym{frr}{FRR}{Free Range Routing}
\newacronym{ups}{UPS}{Uninterruptible Power Supply}

\newacronym{ntia}{NTIA}{National Telecommunications and Information Administration}
\newacronym{pii}{PII}{Personal and Identifiable Information}
\newacronym{irb}{IRB}{Institutional Review Board}
\newacronym{doi}{DOI}{Digital Object Identifier}

\newacronym{sdo}{SDO}{Standard-Development Organization}
\newacronym{ose}{OSE}{Open Source Ecosystem}
\newacronym{osc}{OSC}{O-RAN Software Community}
\newacronym{dop}{DOP}{Director of Operations}
\newacronym{pm}{PM}{Program Manager}
\newacronym{excom}{EXCOM}{Executive Committee}
\newacronym{iiot}{IIoT}{Industrial \gls{iot}}
\newacronym{lf}{LF}{Linux Foundation}

\newacronym{wiot}{WIoT}{Institute for the Wireless Internet of Things}
\newacronym{rl}{RL}{Reinforcement Learning}
\newacronym{drl}{DRL}{Deep Reinforcement Learning}

\newacronym{nofo}{NOFO}{Notice of Funding Opportunity}

\newacronym{onr}{ONR}{Office of Naval Research}
\newacronym{afosr}{AFOSR}{Air Force Office of Scientific Research}
\newacronym{afrl}{AFRL}{Air Force Research Laboratory}
\newacronym{arl}{ARL}{Army Research Laboratory}

\newacronym{arc}{ARC}{Aerial Research Cloud}
\newacronym{cast}{CaST}{Channel emulation scenario generator and Sounder Toolchain}

\newacronym{mno}{MNO}{Mobile Network Operator}
\newacronym{ct}{CT}{Continuous Testing}
\newacronym{oci}{OCI}{Open Container Initiative}

\newacronym{xai}{XAI}{Explainable AI}
\newacronym{esc}{ESC}{Environmental Sensing Capability}
\newacronym{sas}{SAS}{Spectrum Access System}

\newacronym{rem}{REM}{Random Ensemble Mixture}
\newacronym{ns3}{ns-3}{Network Simulator 3}
\tikzstyle{startstop} = [rectangle, rounded corners, minimum width=2cm, minimum height=0.5cm,text centered, draw=black]
\tikzstyle{io} = [trapezium, trapezium left angle=70, trapezium right angle=110, minimum width=3cm, minimum height=1cm, text centered, draw=black]
\tikzstyle{process} = [rectangle, minimum width=2cm, minimum height=0.5cm, text centered, draw=black, alignb=center]
\tikzstyle{decision} = [ellipse, minimum width=2cm, minimum height=1cm, text centered, draw=black]
\tikzstyle{arrow} = [thick,<->,>=stealth]
\tikzstyle{line} = [thick,>=stealth]
\tikzstyle{darrow} = [thick,<->,>=stealth,dashed]
\tikzstyle{sarrow} = [thick,->,>=stealth]
\tikzstyle{larrow} = [line width=0.3mm,dashdotted,->,>=stealth]
\tikzstyle{llarrow} = [line width=0.1mm,->,>=stealth]

\makeatletter
\def\grd@save@target#1{%
  \def\grd@target{#1}}
\def\grd@save@start#1{%
  \def\grd@start{#1}}
\tikzset{
  grid with coordinates/.style={
    to path={%
      \pgfextra{%
        \edef\grd@@target{(\tikztotarget)}%
        \tikz@scan@one@point\grd@save@target\grd@@target\relax
        \edef\grd@@start{(\tikztostart)}%
        \tikz@scan@one@point\grd@save@start\grd@@start\relax
        \draw[minor help lines] (\tikztostart) grid (\tikztotarget);
        \draw[major help lines] (\tikztostart) grid (\tikztotarget);
        \grd@start
        \pgfmathsetmacro{\grd@xa}{\the\pgf@x/1cm}
        \pgfmathsetmacro{\grd@ya}{\the\pgf@y/1cm}
        \grd@target
        \pgfmathsetmacro{\grd@xb}{\the\pgf@x/1cm}
        \pgfmathsetmacro{\grd@yb}{\the\pgf@y/1cm}
        \pgfmathsetmacro{\grd@xc}{\grd@xa + \pgfkeysvalueof{/tikz/grid with coordinates/major step x}}
        \pgfmathsetmacro{\grd@yc}{\grd@ya + \pgfkeysvalueof{/tikz/grid with coordinates/major step y}}
        \foreach \x in {\grd@xa,\grd@xc,...,\grd@xb}
        \node[anchor=north] at (\x,\grd@ya) {\pgfmathprintnumber{\x}};
        \foreach \y in {\grd@ya,\grd@yc,...,\grd@yb}
        \node[anchor=east] at (\grd@xa,\y) {\pgfmathprintnumber{\y}};
      }
    }
  },
  minor help lines/.style={
    help lines,
    gray,
    line cap =round,
    xstep=\pgfkeysvalueof{/tikz/grid with coordinates/minor step x},
    ystep=\pgfkeysvalueof{/tikz/grid with coordinates/minor step y}
  },
  major help lines/.style={
    help lines,
    line cap =round,
    line width=\pgfkeysvalueof{/tikz/grid with coordinates/major line width},
    xstep=\pgfkeysvalueof{/tikz/grid with coordinates/major step x},
    ystep=\pgfkeysvalueof{/tikz/grid with coordinates/major step y}
  },
  grid with coordinates/.cd,
  minor step x/.initial=.5,
  minor step y/.initial=.2,
  major step x/.initial=1,
  major step y/.initial=1,
  major line width/.initial=1pt,
}
\makeatother

\usepackage{dblfloatfix}
\usepackage{float}

\usepackage{xspace}
\newcommand{\oran}{O-RAN\xspace}
\newcommand{\ran}{\gls{ran}\xspace}

\newcommand{\ngrg}{\gls{ngrg}\xspace}

\begin{document}

\title{LibIQ: Toward Real-Time Spectrum Classification in O-RAN dApps}

\author{\IEEEauthorblockN{Filippo Olimpieri\IEEEauthorrefmark{1}, Noemi Giustini\IEEEauthorrefmark{1}, Andrea Lacava\IEEEauthorrefmark{1}\IEEEauthorrefmark{2}, Salvatore D'Oro\IEEEauthorrefmark{2}, Tommaso Melodia\IEEEauthorrefmark{2}, Francesca Cuomo\IEEEauthorrefmark{1}}
\IEEEauthorblockN{\IEEEauthorrefmark{1}Sapienza University of Rome, Rome, Italy\\
\IEEEauthorrefmark{2}Institute for the Wireless Internet of Things, Northeastern University, Boston, MA, USA\\
Email: \{{olimpieri.1933529,\,giustini.1933541\}@studenti.uniroma1.it},\\
Email: \{{lacava.a,\,s.doro,\,t.melodia\}@northeastern.edu},\,\{{francesca.cuomo\}@uniroma1.it}
}
\thanks{Filippo Olimpieri and Noemi Giustini contributed equally to this work.
This article is based upon work partially supported by OUSD(R\&E) through Army Research Laboratory Cooperative Agreement Number W911NF-24-2-0065. The views and conclusions contained in this document are those of the authors and should not be interpreted as representing the official policies, either expressed or implied, of the Army Research Laboratory or the U.S. Government. The U.S. Government is authorized to reproduce and distribute reprints for Government purposes notwithstanding any copyright notation herein.
The work was also partially supported by SERICS (PE00000014) 5GSec project, CUP B53C22003990006, under the MUR National Recovery and Resilience Plan funded by the European Union - NextGenerationEU, and by the U.S.\ National Science Foundation under CNS-1925601.} % <- this stops a space
}
\maketitle

\begin{abstract}
The O-RAN architecture is transforming cellular networks by adopting \gls{ran} softwarization and disaggregation concepts to enable data-driven monitoring and control of the network.
Such management is enabled by \glspl{ric}, which facilitate near-real-time and non-real-time network control through xApps and rApps.
However, they face limitations, including latency overhead in data exchange between the \gls{ran} and \gls{ric}, restricting real-time monitoring, and the inability to access user plain data due to privacy and security constraints, hindering use cases like beamforming and spectrum classification.
To address these limitations, several architectural proposals have been made, including dApps, i.e., applications deployed within the \gls{ran} unit that enable real-time inference, control and \gls{rf} spectrum analysis.
In this paper, we leverage the dApps concept to enable real-time \gls{rf} spectrum classification with LibIQ, a novel library for \gls{rf} signals that facilitates efficient spectrum monitoring and signal classification by providing functionalities to read I/Q samples as time-series, create datasets and visualize time-series data through plots and spectrograms.
Thanks to LibIQ, I/Q samples can be efficiently processed to detect external \gls{rf} signals, which are subsequently classified using a \gls{cnn} inside the library.
To achieve accurate spectrum analysis, we created an extensive dataset of time-series-based I/Q samples, representing distinct signal types captured using a custom dApp running on a  \gls{5g} deployment over the Colosseum network emulator and an \gls{ota} testbed.
We evaluate our model by deploying LibIQ in heterogeneous scenarios with varying center frequencies, time windows, and external \gls{rf} signals.
In real-time analysis, the model classifies the processed I/Q samples, achieving an average accuracy of approximately 97.8\% in identifying signal types across all scenarios.
% We pledge to release both LibIQ and the dataset created as a publicly available framework upon acceptance.
\end{abstract}

\glsresetall


\section{Introduction}
\label{sec:intro}

The Open \gls{ran} paradigm is gaining momentum in the cellular networks thanks to its open and modular approach that promotes interoperability and innovation by disaggregating the \gls{ran} functionalities over white-box hardware and different software components~\cite{polese2023understanding}. 
In the O-\gls{ran} architecture, network control is realized through closed control loops—continuous cycles of data collection and control actions, in which specialized applications can adapt the network to evolving conditions. 
However, in the current standard defined by the O-\gls{ran} ALLIANCE, two key limitations persist: the lack of attention for the user-plane and physical-layer data such as I/Q samples, and the support for sub-10\,ms control loops, ruling out the real-time feedback needed for advanced low-latency use cases, such as beamforming, spectrum sharing, and more~\cite{dappsOranReport}.
The introduction of real-time control loops will enable direct action at the lower layers of the \gls{ran}, allowing adaptive responses to radio conditions while minimizing delays.   
Moreover, a direct access to I/Q samples enables classification systems to detect anomalous signals and interference, enhancing attack detection, dynamic resource management, and spectrum sharing to protect network integrity of the \gls{5g} and beyond cellular networks.   
To address these challenges, dApps~\cite{lacava2025dapps} are introduced as software components deployed alongside \gls{cu} and \gls{du}, where user-plane and I/Q data is directly accessible, enabling \gls{ai}/\gls{ml} routines to operate in real-time within the \gls{ran}.
However, the challenges of processing large I/Q data in real-time remains, requiring specialized \gls{ai}/\gls{ml} models optimized for edge inference~\cite{lacava2025dapps}.

In this paper, we introduce LibIQ\footnote{LibIQ is publicly available at \url{https://github.com/wineslab/lib-iq}.}, a novel library designed to enhance the analysis, manipulation, and labeling of time-series-based I/Q samples, thereby enabling spectrum monitoring and signal classification within dApps.
The library supports both frequency-domain and time-domain analysis, offering functionalities such as plots and spectrogram generation, \gls{fft}  and \gls{psd} calculation.
We embed our library in a dApp to perform spectrum sensing by accessing I/Q samples retrieved from unscheduled \gls{ofdm} symbols to classify in real-time the presence of external \gls{rf} signals and transmission technologies operating in the same spectral space of the \gls{ran} where the dApp is deployed.
Such classification is achieved by LibIQ thanks to an internal \gls{cnn} that is able to detect \glspl{rfi} by analyzing time series based I/Q samples in the frequency domain. 
Such I/Q samples were collected in an extensive data collection campaign with more than 35,200\,time series across six different signal types and four center frequencies, varying the time window to analyze temporal patterns.
Furthermore, we extend the \gls{gui} currently available in the dApp framework by incorporating the output of our model, i.e., the classification labels assigned to the I/Q samples processed in real-time.

To validate the effectiveness of our approach, we conduct experiments in two different \gls{sdr}-based environments: the Colosseum network emulator~\cite{polese2024colosseum}, and an \gls{ota} testbed.
The main contributions of this work are:  
\begin{itemize}
    \item \textit{LibIQ}, a software library for the parsing, manipulation, visualization, and classification of time series-based I/Q samples;
    \item The implementation of a high-performance \gls{cnn} time-series classifier, achieving an average classification accuracy of approximately 97.8\% across diverse \gls{rf} conditions, varying center frequencies, and different time windows;
    \item The first integration of a spectrum classification process into the dApp framework, allowing experiments while respecting the real-time constraints.
\end{itemize}

The remainder of this paper is organized as follows: Section~\ref{sec:soa} reviews existing literature on the O-\gls{ran} architecture and \gls{rf} spectrum classification; Section~\ref{sec:system} presents the functionalities of LibIQ and its integration in the dApp framework;
Section~\ref{sec:experiments} details the experimental pipeline that leverages LibIQ; Section~\ref{sec:results} discusses the results of training and testing of the \gls{cnn}, and Section~\ref{sec:conclusions} summarizes key findings and outlines future research directions.

\section{Related works}
\label{sec:soa}
This section briefly introduces the O-\gls{ran} architecture, the concept of dApps and reviews the various approaches proposed in the literature for spectrum classification.

\subsection{O-RAN and dApps}

The O-\gls{ran} paradigm is a new architectural concept for cellular networks based on disaggregation and programmability of the \gls{ran} functions for enabling a multi-vendor, data-driven approach~\cite{polese2023understanding}.
O-\gls{ran} disaggregates the classic \gls{gnb} into \gls{cu}, \gls{du}, and \gls{ru}, introducing the \glspl{ric} as the entity responsible for the control and managements of the units through open interfaces.
Thanks to this approach, the \glspl{ric} become centralized abstractions of the networks that can manage near-real-time and non-real-time tasks through xApps and rApps.
However, at the current time, there is no standardized way to access real-time, i.e., sub-10\,ms timescales, parameters in the \gls{ran}, therefore no real-time control is available.
To address these issues, a recent research report~\cite{dappsOranReport} from the \oran \ngrg explores the concept of dApps. 
dApps are microservices co-located with \gls{cu} and \gls{du} that enable the real-time exposure and processing of \ran data that otherwise would not be available to the \glspl{ric} due to latency or privacy constraints~\cite{lacava2025dapps}.
In this paper, we leverage dApps by embedding our library's \gls{cnn} to classify \gls{rf} signal types based on processed user-plane data. 
To the best of our knowledge, this is the first work to apply the dApp framework to perform real-time spectrum classification within the O-\gls{ran} context.

%\begin{figure*}[t]
%    \centering
%    \includegraphics[width=0.9\textwidth]{figures/LibIQPipeline.pdf}
%    \caption{In depth overview of the operations performed by LibIQ.}
%    \label{fig:LibIQpipeline}
%\end{figure*}

\subsection{Spectrum Classification with Deep Learning}

Recent research in spectrum sensing and signal classification has explored a range of deep learning techniques to enhance accuracy, efficiency, and adaptability across diverse conditions. Several works have leveraged \glspl{cnn} to process spectral data, each with distinct methodologies and results.

For instance,~\cite{zheng2020spectrum} applies a residual \gls{cnn} to normalized power spectrum data for binary spectrum sensing, effectively distinguishing between signal-plus-noise and noise-only scenarios.
Meanwhile,~\cite{zhang2021signal} integrates a \gls{cnn} with an \gls{lstm} to analyze STFT-based spectrograms, achieving high accuracy in identifying wireless technologies such as Wi-Fi and \gls{5g}.
Other works focus on \gls{cnn}-based architectures specifically optimized for hardware efficiency.
The authors in~\cite{lees2019deep} employ a three-layer \gls{cnn} designed to minimize computational complexity, making it suitable for embedded systems with constrained processing power. 
In contrast,~\cite{amiciradiodl} transforms I/Q samples into image representations and applies the YOLOv8x object detection model, achieving 77.78\% accuracy on a small dataset.
Similarly,~\cite{hauser2017signal} finds that frequency-domain I/Q samples provide better inputs for \gls{cnn}-based modulation classification compared to time-domain representations, underscoring the importance of input feature selection.

To improve latency and efficiency, alternative \gls{cnn} architectures avoid spectrogram conversion.
Spectrum Stitching~\cite{uvaydov2024stitching} uses a U-Net with a non-local block for direct I/Q processing, reducing computational overhead and speeding up inference. 
While this method eliminates feature extraction, it requires extensive training to generalize across signal variations, balancing accuracy and efficiency.
In the context of O-RAN-based implementations, ChARM~\cite{9796985} provides a real-time, data-driven framework for dynamic spectrum sharing in O-RAN networks, enabling spectrum sensing, interference detection. 
However, its integration with \oran remains unclear, and its classifier handles a limited set of \gls{rf} signals.

Our approach differs from previous works in multiple aspects. 
Unlike spectrogram-based classification (\cite{zhang2021signal},~\cite{amiciradiodl}) or hardware-efficient \glspl{cnn} (\cite{lees2019deep}), we process frequency-domain I/Q samples directly using a \gls{cnn} trained on time-series data that performs multinomial classification, in contrast with the binary one of~\cite{zheng2020spectrum}. 
Before classification, we apply an Energy Peak Detector to isolate relevant signals and reduce dimensionality, optimizing both performance and efficiency.
In terms of results, our method achieves 97.8\% classification accuracy, significantly outperforming image-based classification (\cite{amiciradiodl}, 77.78\%) and offering superior real-time adaptability compared to methods based on static spectrograms (\cite{hauser2017signal}, approx 90\%). 
Furthermore, our spectrum classification model is the only one that has been tested within a real dApp framework, interacting directly with an operating \gls{gnb}, unlike previous works that primarily rely on offline datasets or simulated environments or substantial modification of the \oran architecture~\cite{9796985}.

\section{System Overview}
\label{sec:system}

This section provides an overview of LibIQ and its integration into a dApp to enable real-time spectrum analysis.
LibIQ is implemented in C++ and wrapped in Python using \gls{swig} and has four main components. The order in which the library's packages and classes are presented in this chapter reflects the typical workflow a user can follow to analyze and classify its data.

\subsection{\texttt{Analyzer}: Time series analysis and manipulation}

LibIQ provides a package called \texttt{Analyzer} that includes a suite of methods enabling users to manipulate and analyze I/Q sample time series. 
Starting from binary data, LibIQ can parse, extract and select the real or the imaginary component, or both, of the I/Q samples through a set of software functions.
Moreover, LibIQ exposes utility functions such as the \gls{fft} and \gls{psd} to analyze I/Q samples in the frequency domain. 

\subsection{\texttt{Plotter}: Time series visualization}

The \texttt{Plotter} package provides methods for visualize I/Q data. The \texttt{scatterplot()} function allows users to generate scatterplots, which are essential for analyze the distribution and characteristics of signal data. 
These plots can represent the I/Q samples single components or their magnitude and phase.
Additionally, it supports spectrogram generation through \texttt{spectrogram()} function. 
This is achieved by dividing the time series into multiple equal parts (windows) and applying the \gls{fft} to each of them. 
Users can customize window size and overlap to adjust the level of detail in the spectrogram.
Smaller windows with more overlap capture subtle signal variations, otherwise larger windows with less overlap offer a broader view, highlighting the signal’s long-term behavior.

\subsection{\texttt{Preprocessor}: Time Series Preprocessing}

The \texttt{Preprocessor} package handles data preprocessing phase, ensuring the dataset is correctly formatted and optimized for training.
Such parsing is implemented through two key functions: \texttt{create\_dataset\_from\_bin()}, which creates a dataset from a binary file, and \texttt{create\_dataset\_from\_csv()}, which processes a CSV file containing time-series-based I/Q samples. 
Finally, the \texttt{energy\_peak\_detector()} function acts as an Energy Peak Detector that is able to extract and isolate the \gls{rfi} signal and to ensure that the \gls{cnn} model is independent from the \gls{rfi} source center frequency.

\subsection{\texttt{Classifier}: Time Series Classification}

The \texttt{Classifier} class is responsible for training and testing a \gls{cnn} model designed for \gls{rf} signal classification. 
The training process, managed by \texttt{cnn\_train()}, involves feeding the model with time series made of I/Q samples, leveraging features such as the real and imaginary parts, magnitude, and phase.
The length of each time series, i.e., the time window, can change through different trained models as shown more in details in Section~\ref{sec:experiments}.
Once trained, the \gls{cnn} model can be used for classification via the \texttt{cnn\_test()} function, which takes a pre-trained model and predicts the appropriate label for the input time series. 
% Thanks to this mechanism, the \gls{ran} iUsers simply provide one or more time series samples and LibIQ automatically assigns the corresponding labels.

\begin{figure*}[ht]
    \centering
    \includegraphics[width=0.95\textwidth]{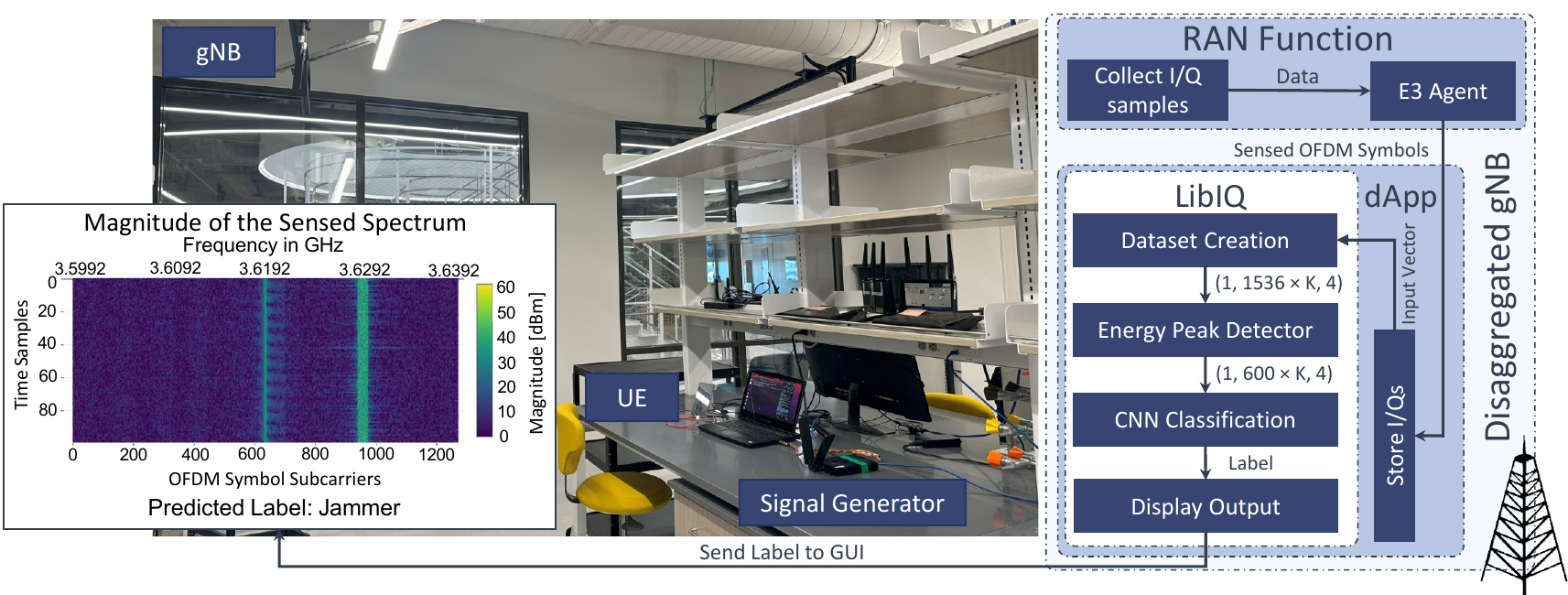}
    \caption{LibIQ deployment inside an O-RAN dApps.}
    \label{fig:ExperimentPipeline}
\end{figure*}

\subsection{Integration of LibIQ on dApp}

As described in Section~\ref{sec:soa}, the O-\gls{ran} ALLIANCE has introduced xApps and rApps to enable near-real-time and non-real-time closed control loops in the \gls{ran}, leaving a gap for the real-time domain.
dApps can perform real-time analysis and interference detection without requiring any \gls{ric}, thus reducing latency.
In this work, we extend the spectrum analysis capabilities of the dApps using LibIQ to create time-series data from the dApp measurements. 
As shown in \figurename~\ref{fig:ExperimentPipeline}, the dApp embeds LibIQ in its real-time control loop to feed the LibIQ pipeline with the I/Q samples.
Then LibIQ applies the Energy Peak Detector, calculates the features and classifies such peaks with the \gls{cnn}.
Once the model has assigned a label to the time series, the result is returned to the dApp, which can propagate it to the \gls{ric} or, as we did in this work, display it on the dApp \gls{gui} beside the “Predicted Label” field in the waterfall plot illustrated in \figurename~\ref{fig:ExperimentPipeline}.
This dApp-based processing approach offers significant advantages by enabling execution directly at the \gls{du}/\gls{cu} without relying on the \gls{ric}.
It enhances efficiency by processing user-plane data locally, eliminating the need for high-bandwidth data transfers over external interfaces. 
The integration of LibIQ enhances network reliability and spectrum analysis by combining real-time processing with \gls{ai}-driven classification.

\section{Real-Time Spectrum Classification}
\label{sec:experiments}

In this section, we introduce the experimental pipeline used to accurately classify I/Q data. Our pipeline is designed to efficiently manage every stage of the process, including analysis, visualization, and classification of \gls{rfi} signals.

\subsection{Dataset collection and preprocessing}
\label{sec:DatasetCollection}

We leverage and extend the dApp framework of~\cite{lacava2025dapps} to access and collect the I/Q samples in real-time.
The dApp is designed with a sensing periodicity of approximately 8\,ms, meaning it periodically provides data to LibIQ for labeling.
Once collected, the I/Q samples are organized into complex input vectors, each containing real and imaginary components. 
Each vector consists of 1,536 I/Q samples, representing data collected from spectrum sensing. These vectors are stored in a binary file, with each file containing 100 input vectors.

We collected data from two different environments: one in an emulated environment using Colosseum~\cite{polese2024colosseum}, and the other in a real environment using transmissions over the air. 
We use a \gls{5g} \gls{sa} deployment of one \gls{gnb} with \gls{oai}, operating at a center frequency of 3.6192\,GHz with a 40\,MHz bandwidth that onboard our dApp. 
As shown in \figurename~\ref{fig:ExperimentPipeline}, for the \gls{ota} testbed we connect a \gls{cots} One Plus Nord phone to the network to generate downlink traffic even though in this work we focus more on the external than the \gls{ue} performance.
To generate external signals, we employed GNU Radio, the \texttt{siggen} utility of the uhd library for \glspl{sdr}, and the SCOPE framework~\cite{bonati2021scope}, each transmitting different signal types, as shown in Table~\ref{tab:signal_labels}.
The center frequency of the transmitted signals varied among 3.6042\,GHz, 3.6142\,GHz, 3.6242\,GHz, and 3.6342\,GHz, while the sampling rate remained fixed at 1\,MHz throughout all transmissions.

\begin{figure}[H]
    \centering
    \begin{subfigure}{0.45\linewidth}
        \centering
        \includegraphics[width=\linewidth]{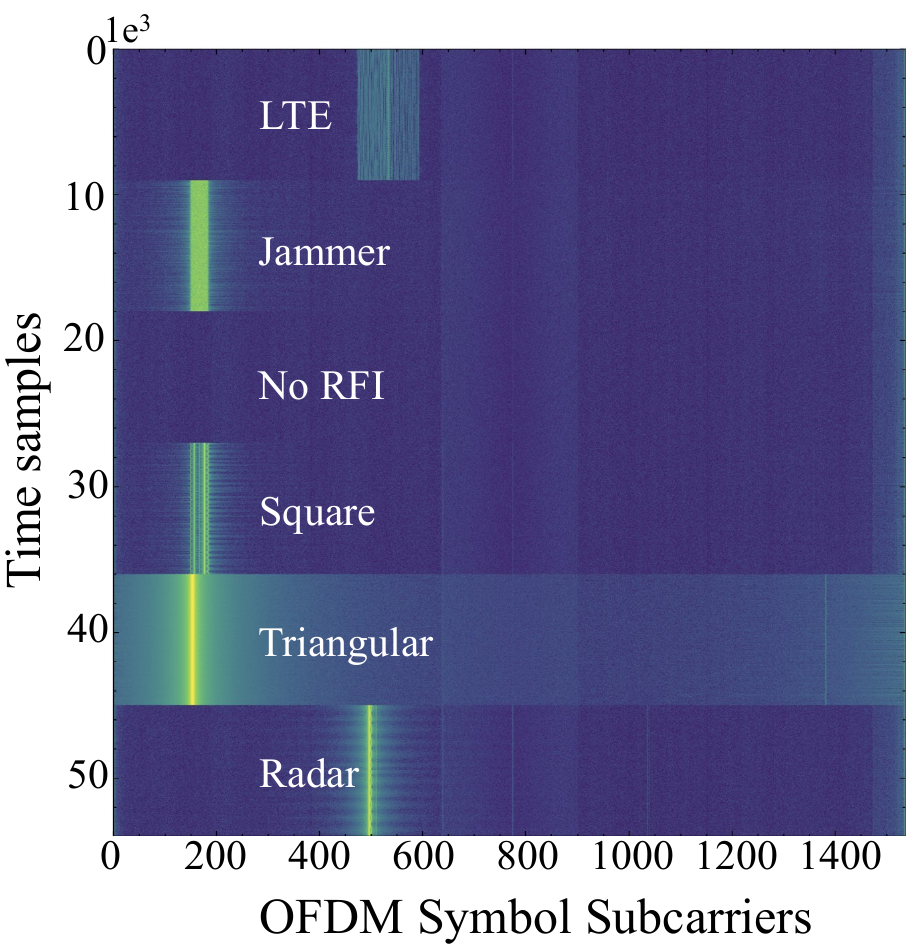}
        \caption{Before Detector.}
        \label{fig:waterfall_Before_ED}
    \end{subfigure}
    \hfill
    \begin{subfigure}{0.52\linewidth}
        \centering
        \includegraphics[width=\linewidth]{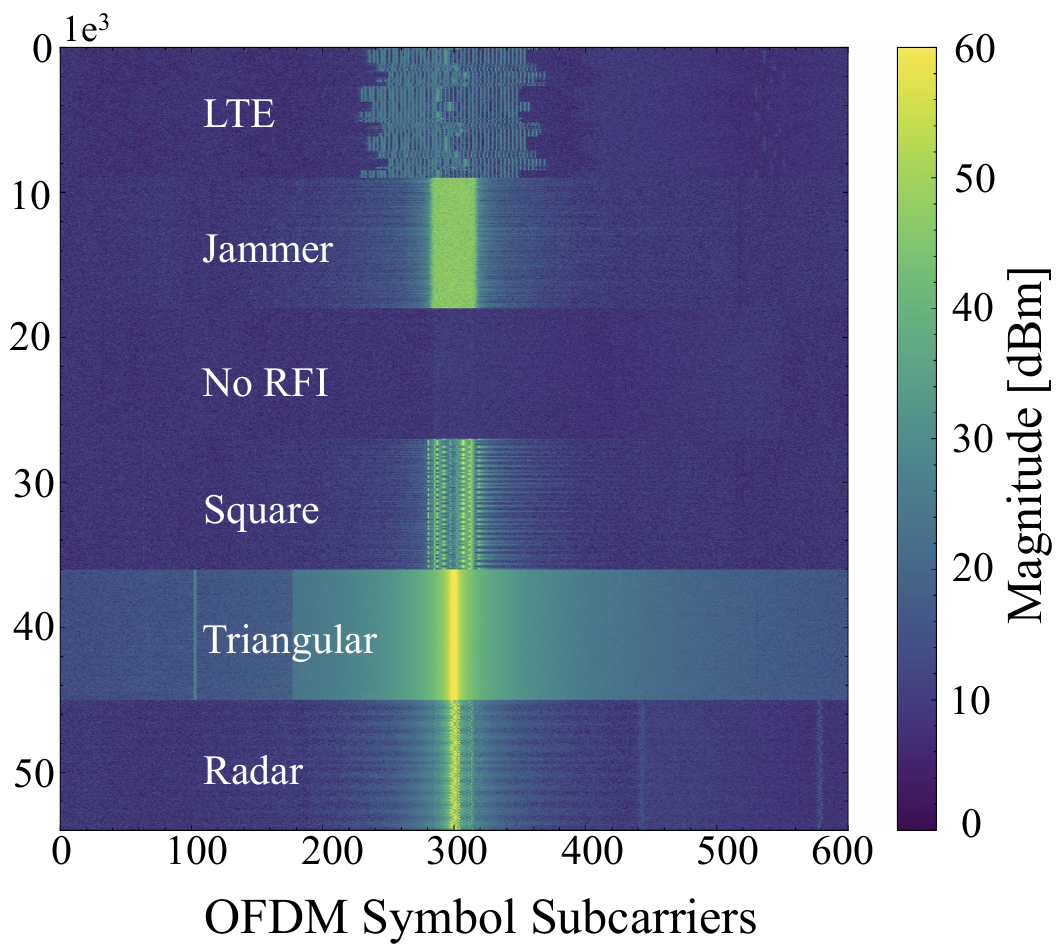}
        \caption{After Detector.}
        \label{fig:waterfall_After_ED}
    \end{subfigure}
    \caption{Comparison of RF data before and after the Energy Peak Detector.}
    \label{fig:waterfall_comparison}
\end{figure}

We organize the data into a structured format with the shape \((N, 1536 \times K, 4)\), where \(N\) represents the total number of time series, while \(K\) is the number of I/Q samples within each time series, which varies depending on the time window. 
In this work, we have tested using time windows of different lengths, i.e., 1, 5, 10, or 15 data samples, to determine how many input vectors are considered for a single prediction.
However, such lengths are not fixed and can be changed and adapted within LibIQ.
A larger time window provides a broader temporal perspective of the signal, capturing more variations over time.   
Lastly, the third dimension \(4\) refers to the number of different features analyzed, i.e., the real part (Q), imaginary part (I), the magnitude, and the phase of the data.
We collected about 800 time series for each \gls{rfi} signal type (LTE, Jammer, No \gls{rfi}, Square, Triangular, and Radar), each environment, and each center frequency (four in total), except for LTE, which was registered only in the Colosseum environment, collecting a total of about 35,200 time series.
The details of these \gls{rfi} signals are provided in Table~\ref{tab:signal_labels} and illustrated in \figurename~\ref{fig:waterfall_comparison}.

\begin{table}[ht]
    \centering
    \setlength{\tabcolsep}{2pt}
    \renewcommand{\arraystretch}{1}
    \begin{tabular}{p{1.2cm} cccc p{1.2cm} p{1.8cm}}
        \toprule
        \textbf{Waveform} & \multicolumn{2}{c}{\textbf{Gain [dB]}} & \multicolumn{2}{c}{\textbf{Amplitude}} & \textbf{Label} & \textbf{Software} \\
        \cmidrule(lr){2-3} \cmidrule(lr){4-5}
        & \textbf{OTA} & \textbf{Colosseum} & \textbf{OTA} & \textbf{Colosseum} &  &  \\
        \midrule
        Sine        & $90$ & $25$ & $1$ & $1$ & Radar      & Siggen \\
        Triangular  & $100$ & $25$ & $1$ & $1$ & Triangular & GNU Radio \\
        Square      & $100$ & $25$ & $1$ & $1$ & Square     & GNU Radio \\
        Uniform     & $80$ & $20$ & $1$ & $1$ & Jammer     & Siggen \\
        LTE         & $/$ & $20$ & $/$ & $0.9$ & LTE & SCOPE~\cite{bonati2021scope} \\
        \bottomrule
    \end{tabular}
    \caption{Waveform types and their assigned labels in the experimental dataset.}
    \label{tab:signal_labels}
\end{table}
Finally, we apply an Energy Peak Detector that identifies the maximum energy peak within a user-defined sliding window. Then, a specified number of samples are extracted from both sides, ensuring that the most relevant portion of the signal is preserved. 
After applying the Energy Peak Detector, the data are reshaped into (\( N, J \times K, 4 \)), where \( J \) is the length of the detector window, set to 600 I/Q samples in our experiments.
Figure~\ref{fig:waterfall_Before_ED} shows the raw input before detection, where the full 1536-sample sequence contains noise and irrelevant data. In contrast, Figure~\ref{fig:waterfall_After_ED} shows how the detector isolates a shorter segment of samples, centered on the energy peak, that captures the most relevant portion of each RF signal.
This reduction in input size decreases computational cost by avoiding analysis of the full spectrum. More importantly, by focusing only on the informative part, it improves the accuracy and generalization of the \gls{cnn} models while reducing the risk of overfitting.

The dataset collected was used to train and evaluate the performance of the \gls{cnn}, which will be further discussed in Section~\ref{sec:results}. 
Specifically, we used two center frequencies for training, while the remaining frequencies were reserved for testing to assess the model's generalization capability.

\subsection{Data-Driven Spectrum Classification}
\label{subsec:LabelingCNN}

\begin{figure}[H]
    \centering
    \includegraphics[width=1\columnwidth]{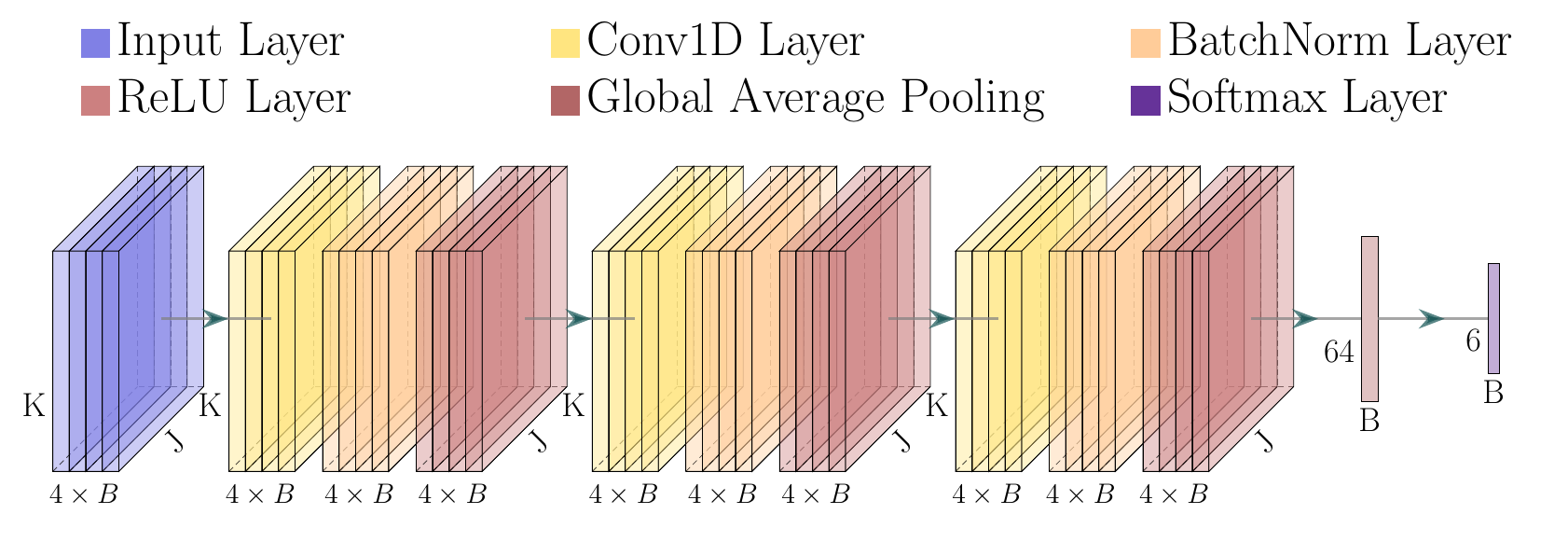}
    \caption{\gls{cnn} Architecture for signal classification.}
    \label{fig:CNN_architecture}
\end{figure}

The \gls{cnn} in our library, represented in \figurename~\ref{fig:CNN_architecture}, is designed to label time-series of I/Q samples based on the transmission technologies of the signals.
The model processes time-series-based I/Q samples with an input shape of (32, \( J \times K \), 4), i.e.,  the batch size, the total number of I/Q samples \( J \times K \) and and the number of distinct features. \( J \) denotes the length of each input vector after applying the energy detector and \( K \) refers to the number of I/Q samples within each time series, that depends on the time window.

The model begins with an input layer that receives the time-series data. 
Then, the first convolutional layer applies 64 one-dimensional filters along the time axis with padding, producing an output of dimension equal to (32, \( J \times K \), 64). 
This is followed by batch normalization that normalizes activations between channels and a ReLU activation function that improves the model's ability to capture non-linear patterns.

This pattern is repeated three times to allow the \gls{cnn} to progressively learn more complex features. 
After this feature extraction process, a global average pooling layer condenses the learned information into a fixed-length feature vector with shape (32, 64). 
After this phase, the extracted features are passed through a fully connected layer, transforming them into a representation with shape (32, 6) that captures higher-level abstractions of the time-series data.  
Finally, the model's output layer is made of a Softmax activation layer, which maps the 6-dimensional representation to a number of output units equal to the number of classes, where each output unit represents the probability of a specific class.  

During the training phase, which lasts 10 epochs, the model learns to differentiate between signal types by extracting relevant features from the dataset previously discussed in Section~\ref{sec:DatasetCollection}.  
An epoch consists of a complete pass through all the samples in the training set of dimension (\( N \), \( J \times K \), 4) where \( N \) is equal to 17600. 
The choice of \( N \) comes from the dataset's size that contains 35200 time series, with half of the center frequencies allocated for training, and the remaining two are reserved for testing to evaluate the model's generalization ability.
The \gls{cnn} processes the data in batches of 32 samples so, the number of updates performed during a single epoch is \( 17600 / 32 = 550 \).
This means that the model updates its weights 550 times per epoch through backpropagation. 
Over the course of 10 epochs, the total number of updates amounts to \( 550 \times 10 = 5500 \). 

\section{Results}
\label{sec:results}
In this section, we evaluate the results of the \gls{cnn} model and the experiments introduced in this work, summarized in Table~\ref{tab:TW_metrics}.

\begin{table}[h]
    \centering
    \setlength{\tabcolsep}{4pt}
    \renewcommand{\arraystretch}{1}
    \begin{tabular}{p{1cm} p{2.5cm} c c c c}
        \toprule
        & & \multicolumn{4}{c}{Time Windows [\# input vectors]} \\
        \cmidrule(lr){3-6}
        \textbf{Metric} & \textbf{Description} & \textbf{1} & \textbf{5} & \textbf{10} & \textbf{15} \\
        \midrule
        Accuracy  & Ratio of correct predictions to total predictions & 0.979 & 0.987 & 0.971 & 0.976 \\
        Precision & Ratio of true positives to all predicted positives & 0.982 & 0.988 & 0.975 & 0.979 \\
        Recall    & Ratio of true positives to all actual positives & 0.980 & 0.987 & 0.971 & 0.976 \\
        F1 Score  & Harmonic mean of precision and recall & 0.980 & 0.972 & 0.971 & 0.976 \\
        Latency [ms] & Time required for a single prediction & 
        \makecell{$2.23$ \\ $\pm 0.48$} & 
        \makecell{$1.13$ \\ $\pm 0.04$} & 
        \makecell{$0.92$ \\ $\pm 0.11$} & 
        \makecell{$0.90$ \\ $\pm 0.05$} \\
        \bottomrule
    \end{tabular}
    \caption{Performance metrics over different time windows.}
    \label{tab:TW_metrics}
\end{table}

We use multiple performance indicators to provide a comprehensive evaluation of the model’s effectiveness. Accuracy measures overall correctness, peaking at 98.7\% and never dropping below 97.1\%. Precision ranges from 97.5\% to 98.8\% ensuring the model makes correct predictions, and minimizes false positives. Instead, Recall confirms its ability to detect true positives (mean 97.6\%). Furthermore, the F1 score remains above 97.1\%, confirming a well-balanced trade-off between capturing positive instances and preventing misclassifications.

To evaluate model latency, we ran six three-minute experiments per time window size, computing mean latency and z-scores.
As shown in Table~\ref{tab:TW_metrics}, there is a trade-off between window size and total latency. Larger windows reduce the number of predictions but increase per-prediction time, while smaller windows lead to more frequent, faster predictions, raising overall latency.
For example, with equal experiment duration, a window size of 15 produces a single prediction with minimal buffering but higher inference time, whereas a size of 1 yields 15 quicker predictions, increasing cumulative latency.
However, latency remains consistently below the 10 ms real-time threshold.

\begin{figure}[H]
    \centering

    \begin{subfigure}[b]{0.48\textwidth}
        \centering
        \includegraphics[width=\textwidth]{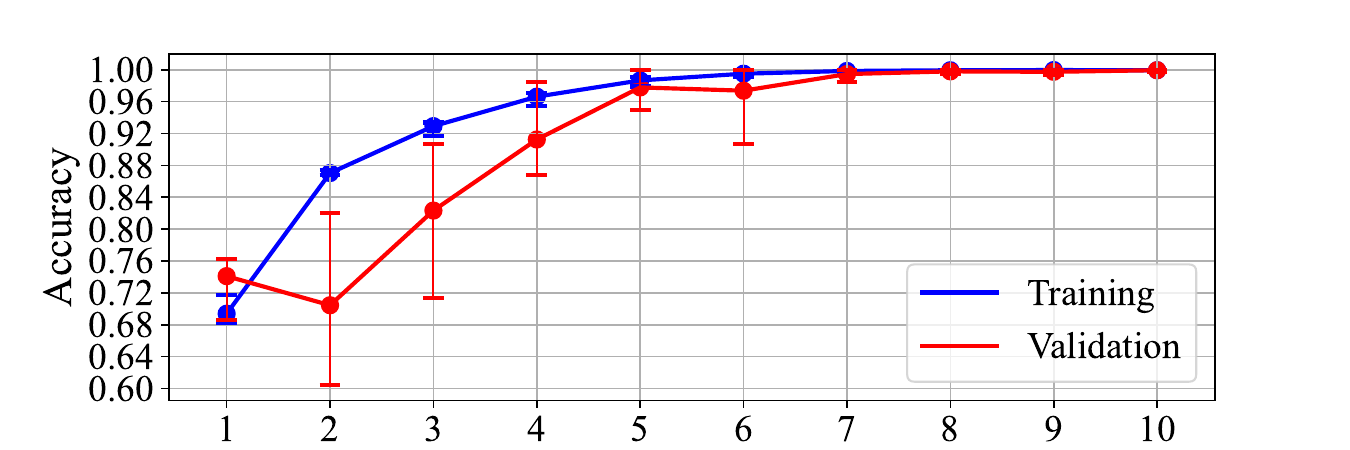}
        \caption{Training and Validation Mean Accuracy.}
        \label{fig:AccuracyCurve}
    \end{subfigure}
    \begin{subfigure}[b]{0.48\textwidth}
        \centering
        \includegraphics[width=\textwidth]{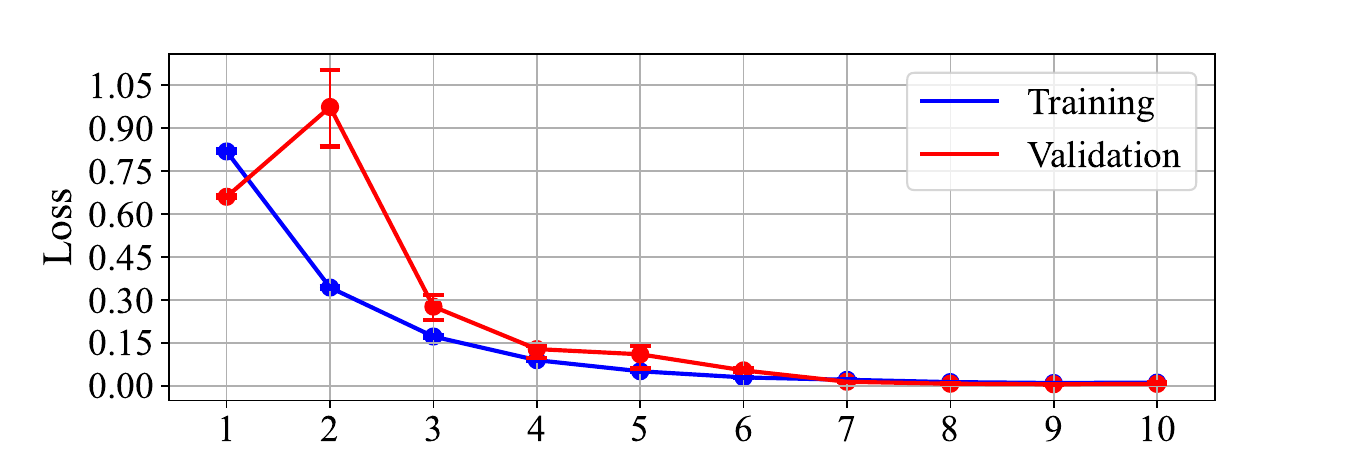}
        \caption{Training and Validation Mean Losses.}
        \label{fig:LossCurve}
    \end{subfigure}

    \caption{Training and validation losses and accuracies over epochs.}
    \label{fig:LossAndAccuracyCurves}
\end{figure}

\begin{figure}[h]
    \centering
    \includegraphics[width=0.34\textwidth]{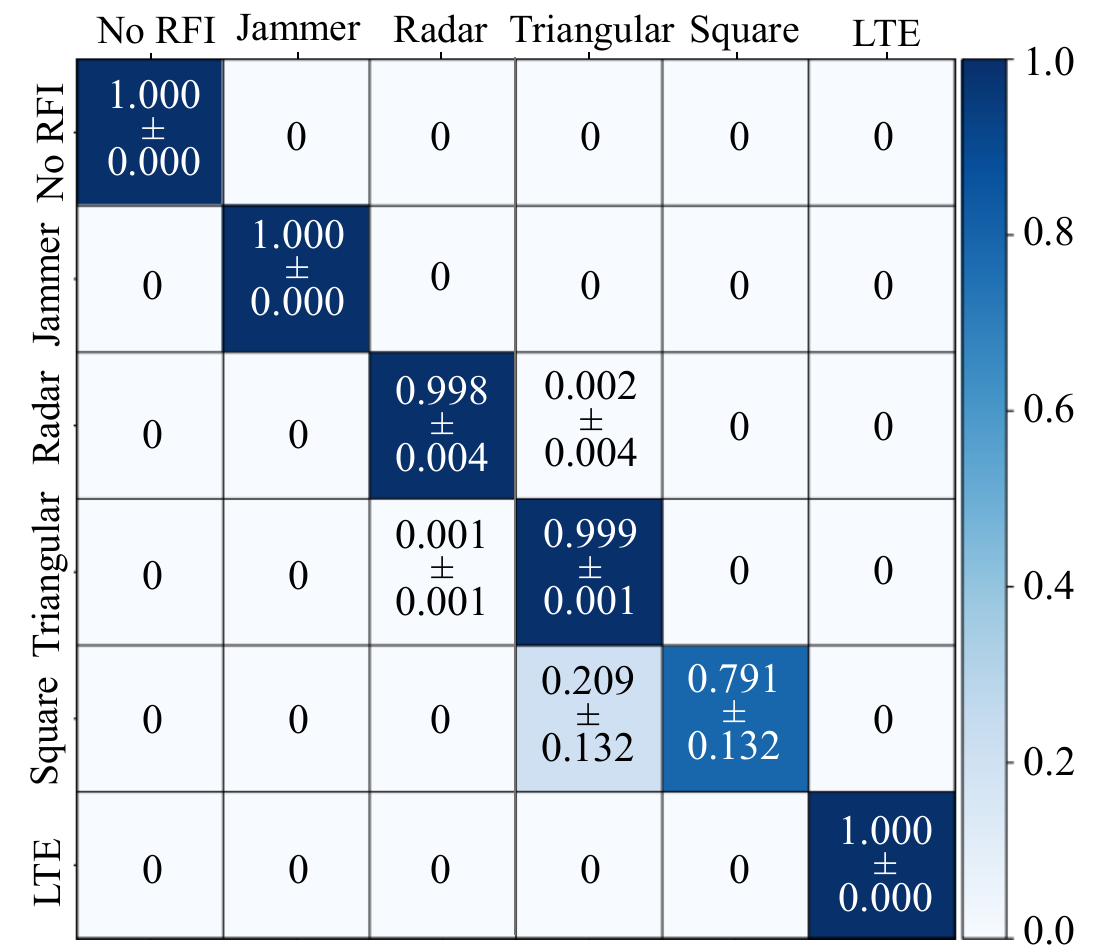}
    \caption{Predicted vs Real Labels for Signal Classification with Z-score.}
    \label{fig:ConfusionMatrix}
\end{figure}

Figures~\ref{fig:AccuracyCurve} and~\ref{fig:LossCurve} illustrate the average training and validation accuracy and loss per epoch, respectively, across multiple experiments. 
% Both curves span 10 training epochs showing the learning phase progression. 
In Figure~\ref{fig:AccuracyCurve}, the accuracy of training and validation increases rapidly and converges to nearly 100\%, indicating fast and effective learning. 
Similarly, Figure~\ref{fig:LossCurve} shows a rapid reduction in both training and validation losses, with the two curves remaining closely aligned throughout the process. This alignment suggests a stable training phase and confirms the absence of overfitting.
\balance
Lastly, Figure~\ref{fig:ConfusionMatrix} reveals a near-perfect classification accuracy with minimal misclassifications, further demonstrating the robustness and effectiveness of the proposed approach.

\section{Conclusions and Future Work}
\label{sec:conclusions}

In this work, we introduce LibIQ, a comprehensive library designed to analyze, manipulate and label time-series-based I/Q samples with high performance.
Thanks to its integration with dApps, LibIQ provides a new way to enable real-time analysis methodologies, increasing awareness and paving the way to full control of the spectrum in the O-\gls{ran} architecture.

An innovative feature of LibIQ is its \gls{cnn} classifier, trained on an extensive and diverse dataset of signals to ensure robust and accurate classification.
To rigorously evaluate its performance, we conducted numerous tests in two \gls{sdr}-based environments, proving the reliability of LibIQ thanks to the combination of the Energy Peak Detector and the \gls{cnn} classifier.
The results demonstrate LibIQ's robustness, as it achieves an average classification accuracy of approximately 97.8\% across different scenarios, center frequencies, signal conditions and environments.

Looking ahead, we plan to enhance the dataset by incorporating a broader range of signal types, emerging technologies, and complex network scenarios to improve system robustness.
Additionally, we aim to integrate anomaly detection algorithms to enable a deeper analysis of spectral, temporal, and behavioral patterns in I/Q traffic.
Finally, a key future direction is to establish a direct connection between the analysis module and the \gls{ric}, enabling automated corrective actions and fostering a more intelligent and adaptive \gls{ran} environment.

\footnotesize
\bibliographystyle{IEEEtran}
\bibliography{biblio}

\end{document}